# Evaluation of Electric and Magnetic Fields Distribution and SAR Induced In 3D Models of Water Containers by Radiofrequency Radiation and Their Relationship to the Non-Thermal Effects of Microwaves

Maher A. A. Abdelsamie[1]*, Russly b Abdul Rahman [1,2,3], Shuhaimi Mustafa1, M.M. Isa[4], Dzulkifly Hashim[1]

[1]Halal Products Research Institute, Universiti Putra Malaysia, Malaysia;
[2]Department of Process & Food Engineering, Faculty of Engineering, Universiti Putra Malaysia, Malaysia;
[3]Faculty of Food Science & Technology, Universiti Putra Malaysia, Malaysia;
[4]Department of Electrical and Electronic Engineering, Faculty of Engineering, Universiti Putra Malaysia, Malaysia.

Corresponding authors:

Maher A. A. Abdelsamie     Email: mabdelaleem2011@gmail.com
Russly b Abdul Rahman     Email: russly@upm.edu.my,
Phone: +603-89430405     Fax : +603-89439745

**Abstract**

The research works on the biological effects of electromagnetic (EM) radiation have increased globally. Computer simulation models were developed to assess exposure of square, rectangular, pyramidal, and cylindrical water containers to microwave radiation at 300, 900, and 2,400 MHz. The development of the models included determination of EM field distribution and the resulting specific absorption rate (SAR) in the stored water. These models employed CST STUDIO SUITE 2014 package to solve EM field equations by applying the finite-difference time-domain (FDTD) method. The effect of frequency, packaging shape, and polarization on SAR induced in water was determined. High electric field and point SAR were obtained over the whole azimuth and elevation angles range in the pyramidal container model. The highest values of SAR were induced in water at the sharp edges of the four water container models. The order of the effect on total SAR and maximum point SAR is cylindrical < square < rectangular < pyramidal model at 300, 900, and 2,400 MHz for both vertical and horizontal polarizations. It can be concluded that the variation in the packaging shape of the containers, polarization, irradiation geometry, and frequency. In addition, the sharp edges of the container models significantly affect the calculation and distribution of electric and magnetic fields and SAR values induced in the stored water, which in turn could cause variations in the non-thermal effects of the electromagnetic fields in the stored water.

## 1. Introduction

The telecommunication revolution of modern society paved the way to a worldwide wireless coverage with necessary infrastructure. The rapidly advancing wireless technologies have fully supported this growing demand for communication and data transfer throughout the world. Unfortunately, the resulting electromagnetic (EM) radiations from mobile base stations, Wi-Fi, and other sources have become a universal phenomenon. Recently, there has been an increase in public concern over the scientific evidence for potential health risks caused by radiofrequency/microwave radiation emissions (RF), which are irradiated from wireless communications. The studies and research regarding the effect of EM waves on human heath reported many serious implications. An important report was published regarding the biological effects and adverse health impacts at exposure levels below current public safety standards, and it documented considerable scientific evidence [1]. There are more than 1800 studies as mentioned in bioinitiative report 2012 [2] confirmed that the radiofrequency radiation at low-intensity exposure emitted from a cell tower, Wi-Fi, wireless laptop, and other sources causes notable biological effects [2]. Specific absorption rate (SAR) is a measure of the rate at which energy is absorbed by the biological material, such as human body or food, when exposed to an EM field, and it is defined as the power absorbed per mass of tissue and has units of watts per kilogram (W/kg) [3]. When biological materials are exposed to EM waves in the radiofrequency range, electric and magnetic fields are induced in the materials, which can be simulated by solving Maxwell's equations for given boundary conditions [4]. The FDTD method has earned worldwide recognition among the researchers because it makes it possible to use spatial and temporal discretization of Maxwell's equations.

Many studies have highlighted the role geometrical shapes of biological materials play in absorbing EM energy [5-9]. For example, a study on the effect of the 900-MHz frequency—which is one of the GSM frequency bands currently being used by GSM mobile phones in many parts of the world on different shapes of the human head models, including flat, ellipsoidal, and spherical shapes—and its relationship to SAR revealed that SAR value was changed by changing the shape of the head model [5]. Geometrical shapes also play a role in the absorption of EM energy in the microwave heating process. For example, the edge overheating phenomenon that occurs during microwave heating is due to the concentration of EM energy at the sharp edges of food products [6]. Water is widely used as the universal solvent in many liquid food products, water-based pharmaceuticals and chemicals. The exposure of water solutions to low-intensity EM radiation in the microwave frequencies of 450 and 2400 MHz was shown to induce nonthermal effects in water solutions, such as effects on the conductivity and viscosity of water caused by effects on the hydrogen bonding network [10; 11]. Recently, a study was conducted on the effects of packaging shape, environmentally abundant electromagnetic fields and storage duration on $^{17}$O NMR and Raman spectra of H2O-NaCl. The study showed that the exposure of H2O-NaCl solution stored inside two shielded and unshielded groups of containers with sharp edges to environmentally abundant EM fields caused variation in the cluster size of water stored in shielded and unshielded containers as well as between the same groups of containers [8]. According to the study, the variations in the cluster size of water could be attributed to the variations in the SAR values induced in water during the storage period. The role that the sharp edges of the containers play in affecting the distribution of electric and magnetic fields and inducing nonthermal effects in H2O-NaCl solutions stored in different packaging shape models was not explained in the earlier study [8].

In this study, the FDTD method was applied to pyramidal, rectangular, square, and cylindrical water container models to investigate how packaging shape, polarization, irradiation geometry, and frequency affect the energy deposition in water in the four container models. In addition, the role of sharp edges of water container models and its relationship to the thermal and nonthermal effects and mechanisms of electromagnetic radiation was also explored, in order to obtain parameters for water solutions storage and packaging process optimization.

2. **Materials and methods**

**2.1. Modeling**

In this study, the CST STUDIO SUITE 2014 package was used to design the three-dimensional (3-D) water container models and to perform SAR simulation by applying the FDTD method. The 3-D models of pyramidal, cylindrical, square, and rectangular containers filled with 100% water was designed as shown in Figure 1. The models consist of two parts: (1) external polymethyl methacrylate (PMMA) layer with 3-mm thickness and (2) water as the inner medium. The properties of PMMA and water are shown in Table 1. The dielectric constants and conductivity of the materials are directly related to the capacitance and the ability of these materials to store energy [12]. This study investigates the effect of packaging shape, plane wave incident angle, polarization, and frequency on the calculation of electric and magnetic fields and SAR absorbed in water in the four water container models. To ensure that the volume of water in the four container models is the same (2.65 L), the dimensions of water in Table 2 were used to solve formulas 1, 2, 3 and 4 for rectangular, pyramidal, square, and cylindrical container models, respectively.

$$V = whl \quad (1)$$

$$V = a^2 \frac{h}{3} \quad (2)$$

$$V = a^2 h \quad (3)$$

$$V = \pi r^2 h \quad (4)$$

**Table 1.** Material Properties Used in the Computational Modeling.

| Material | Relative permittivity $\varepsilon_r$ | electrical conductivity $\sigma$ (S/m) | Mass Density $\rho$ (Kg/m3) |
|---|---|---|---|
| PMMA[a] | 3.6 | 0.02 | 1190 |
| Water | 78 | 1.59 | 1000 |

a. Polymethyl methacrylate (PMMA)

**Table 2.** Dimensions of the four container models.

| Model | Width (mm) | | Depth (mm) | | Radius (mm) | | Height (mm) | |
|---|---|---|---|---|---|---|---|---|
| | Water | PMMA | Water | PMMA | Water | PMMA | Water | PMMA |
| Pyramid-shaped | 230.059 | 242.362 | 230.059 | 242.362 | - | - | 150.227 | 155.10 |
| Cylindrical-shaped | - | - | - | - | 74 | 77 | 154 | 157 |
| Rectangle-shaped | 120 | 126 | 120 | 126 | - | - | 184 | 190 |
| Square-shaped | 144 | 150 | 144 | 150 | - | - | 128 | 134 |

a. Polymethyl methacrylate (PMMA)

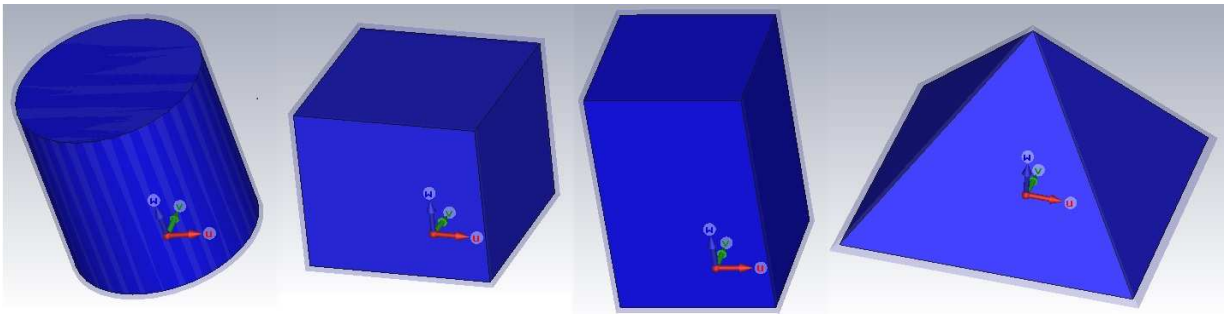

**Figure 1.** 3D models of pyramidal, rectangular, cylindrical and square shaped containers

## 2.2. Exposure setup and FDTD-based SAR computation

The water container models were exposed to free-space 1 V/m plane waves for vertical and horizontal polarizations at 300, 900, and 2,400 MHz. This frequency range covers many applications, such as GSM, various wireless broadband services, and wireless local area network (WLAN). Plane wave excitation was used to excite the surfaces of each model from only one side. The elevation angle (θ) ranged from 0° to 100° with a step width of 20°. Because of the symmetry of the four models, the incident field propagation direction is rotated in 20° steps around the model (0°–180°) as shown in Figure 2 (a-d). Both vertical (V) and horizontal (H) polarizations have been considered, so more than 1,440 numerical simulations have been performed. The EM simulations were performed by applying the FDTD method using the CST STUDIO SUITE 2014 package. The maximum values of electric and magnetic fields, and total SAR for water was evaluated by using template-based post-processing option. The maximum point SAR was calculated by using the exposure geometry of the highest values of total SAR. Total SAR was selected as a parameter of energy absorption in water because total SAR is the power loss density (PLD) of a biological material divided by its mass density, which can be used to differentiate between the four container models in terms of the total energy absorbed in water divided by the mass density of water inside the containers. The maximum point SAR was also selected as a parameter because the maximum point SAR is the maximum point SAR of all grid cells. For each grid cell, its point SAR is calculated by its absorbed power divided by its mass. In-depth investigation of the variability of the SAR on the frequency, incident angle, and polarization of the four models was performed. A perfectly matched layer (PML) was used to represent the absorbing boundary and truncate the computational regions. The number of hexahedral meshing cells was 2,478,175, 2,469,600, 6,795,932, and 3,027,456 in rectangular, square, pyramidal, and cylindrical container models, respectively. Ten cells per wavelength were used to ensure stability.

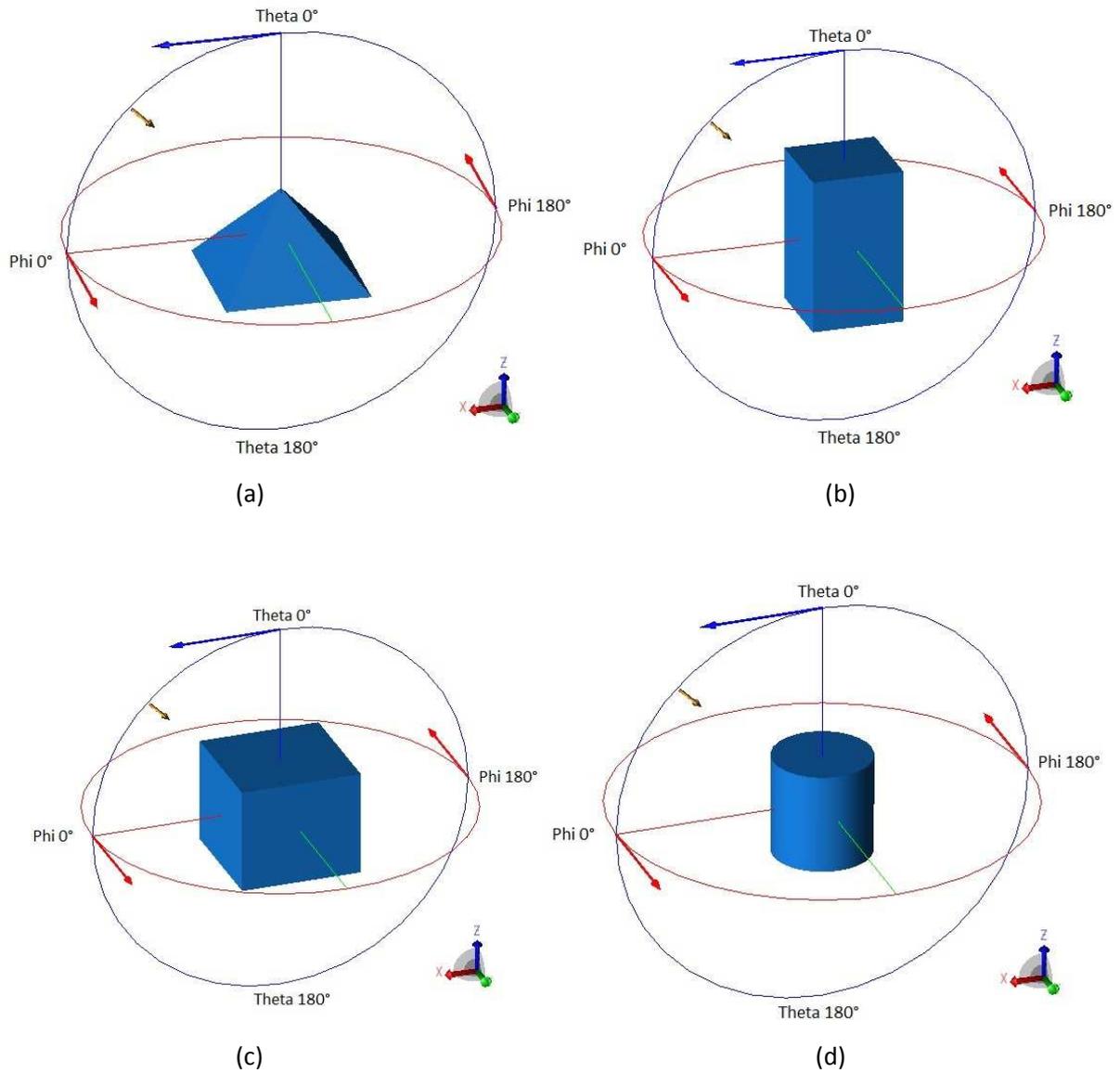

**Figure 2 (a-d).** Definition of the incident angles. Elevation angle (θ) ranged from 0° to 100° with a step width of 20°and azimuth angle (φ) rotates in 20° steps from 0° to 180°.

## 2.3 FDTD governing equations

The general characteristics of EM waves that travel in a specific medium are governed by the following Maxwell's four equations:

$$\nabla \cdot \vec{D} = \rho \tag{1}$$

$$\nabla \cdot \vec{B} = 0 \tag{2}$$

$$\nabla \times \vec{E} = -\mu \frac{\partial \vec{H}}{\partial t} \tag{3}$$

$$\nabla \times \vec{H} = \sigma \vec{E} + \varepsilon \frac{\partial \vec{E}}{\partial t} \tag{4}$$

Where $\vec{D}$, $\vec{B}$, $\vec{E}$, and $\vec{H}$ symbolize electric flux density expressed as (C/m$^2$), Magnetic flux density expressed as (Wb/m2), Electric field intensity expressed as (V/m), Magnetic field intensity expressed as (A/m), respectively. The $\sigma$, $\mu$, $\varepsilon$ and $\rho$ symbolize electric conductivity expressed as (S/m), Permeability expressed as (H/m), Permittivity expressed as (F/m) and Electric charge density per unit volume of the material expressed as (C/m$^3$), respectively [13]. Maxwell's equations 3 and 4 for solving FDTD computational volume in 3-D space considering the electric and magnetic field components can be written as:

$$\nabla \times \vec{E} = \begin{vmatrix} \hat{a}_x & \hat{a}_y & \hat{a}_z \\ \frac{\partial}{\partial x} & \frac{\partial}{\partial y} & \frac{\partial}{\partial z} \\ E_x & E_y & E_z \end{vmatrix} = -\mu \frac{\partial \vec{H}}{\partial t} \tag{5}$$

$$\nabla \times \vec{H} = \begin{vmatrix} \hat{a}_x & \hat{a}_y & \hat{a}_z \\ \frac{\partial}{\partial x} & \frac{\partial}{\partial y} & \frac{\partial}{\partial z} \\ H_x & H_y & H_z \end{vmatrix} = \sigma \vec{E} + \varepsilon \frac{\partial \vec{E}}{\partial t} \tag{6}$$

After considering the components of the EM field equations in the *x*, *y*, and *z* dimensions, it becomes [14]:

$$\frac{\partial H_x}{\partial t} = \frac{1}{\mu}\left(\frac{\partial E_y}{\partial z} - \frac{\partial E_z}{\partial y}\right) \quad (7a)$$

$$\frac{\partial H_y}{\partial t} = \frac{1}{\mu}\left(\frac{\partial E_z}{\partial x} - \frac{\partial E_x}{\partial z}\right) \quad (7b)$$

$$\frac{\partial H_z}{\partial t} = \frac{1}{\mu}\left(\frac{\partial E_x}{\partial y} - \frac{\partial E_y}{\partial x}\right) \quad (7c)$$

$$\frac{\partial E_x}{\partial t} = \frac{1}{E}\left(\frac{\partial H_z}{\partial y} - \frac{\partial H_y}{\partial z} - \sigma E_x\right) \quad (7d)$$

$$\frac{\partial E_y}{\partial t} = \frac{1}{E}\left(\frac{\partial H_x}{\partial z} - \frac{\partial H_z}{\partial x} - \sigma E_y\right) \quad (7e)$$

$$\frac{\partial E_z}{\partial t} = E1\left(\frac{\partial H_y}{\partial x} - \frac{\partial H_x}{\partial y} - \sigma E_z\right) \quad (7f)$$

Equations 5 and 6 do not provide any closed-form solution for irregular or complex geometry. Therefore, the numerical approach is the most appropriate method for solving such equations. The numerical approach creates regular geometries by discretizing the complex or irregular geometry into smaller cubic cells of size ($\Delta x$, $\Delta y$, and $\Delta z$), which is the space incremental at the given time step ($\Delta t$), which is time increment and direction of *x*, *y*, and *z*, respectively. FDTD has been specifically designed to solve EM problems, particularly time-differentiated Maxwell's curl equations (equations 5 and 6). The FDTD method provides a continuous simulation of EM waves in a finite spatial region through a sustained time-marching or time-stepping procedure until the stable field pattern or desired simulation time is achieved [14].

However, the standard unit of a cell, as described by Yee, is a cube ($\Delta x = \Delta y = \Delta z = \delta$) with equal edges. The evaluation of stability factor that determines the stability of numerical simulation makes it possible to use structures with nonequal edges as long as it conforms with the stability factor [15]. The coordinate of Yee's cell can serve as the grid point (*m*, *n*, *p*). The distance of *x*, *y*, and *z* units at any grid point ($x = m\Delta x$, $y = n\Delta y$, and $z = p\Delta z$). The same manner is used for discretizing the time, and by considering a time step $\Delta t$ (i.e., $t = q\Delta t$) at the grid, a specific point in time *q* corresponds to a specific time. Considering the time discretization in FDTD, a partial equation can be represented as the linear equation:

$$\frac{\partial H_i}{\partial t} = \frac{H_i^{q+1/2}(m,n,p) - H_i^{q-1/2}(m,n,p)}{\Delta t}$$

$$\frac{\partial E_i}{\partial t} = \frac{E_i^{q+1}(m,n,p) - H_i^{q}(m,n,p)}{\Delta t}$$

(8)

Considering the position and central differences, a partial differential equation can be represented as the linear equation:

$$\frac{\partial F_i}{\partial x} = \frac{F_i^{q}(m+1,n,p) - F_i^{q}(m,n,p)}{\Delta x}$$

$$\frac{\partial F_i}{\partial y} = \frac{F_i^{q}(m,n+1,p) - F_i^{q}(m,n,p)}{\Delta y}$$

$$\frac{\partial F_i}{\partial z} = \frac{F_i^{q}(m,n,p+1) - F_i^{q}(m,n,p)}{\Delta z}$$

(9)

The updated magnetic field ($H_i^{q+1/2}$) values determine the next electric field on the time-stepping grid while the updated electric field ($E_i^{q+1}$) values determine the next magnetic field on the grid. The continuous updating process of electric and magnetic fields create a time-marching procedure, simulating EM fields as explained by Taflove (2005). The following equations represent the updated fields on the next time step of a time-stepping grid [2] [14]:

$$H_x^{q+\frac{1}{2}}\left(m,n+\frac{1}{2},p+\frac{1}{2}\right) = H_x^{q-\frac{1}{2}}\left(m,n+\frac{1}{2},p+\frac{1}{2}\right) + \left\{\frac{\Delta t}{\mu \Delta z}\left[E_y^{q}\left(m,n+\frac{1}{2},p+1\right) - E_y^{q}\left(m,n+\frac{1}{2},p\right)\right] - \frac{\Delta t}{\mu \Delta y}\left[E_z^{q}(m,n+1,p+\frac{1}{2}) - E_z^{q}\left(m,n,p+\frac{1}{2}\right)\right]\right\} \quad (10)$$

$$H_y^{q+\frac{1}{2}}\left(m+\frac{1}{2},n,p+\frac{1}{2}\right)=H_y^{q-\frac{1}{2}}\left(m+\frac{1}{2},n,p+\frac{1}{2}\right)+\left\{\frac{\Delta t}{\mu\Delta x}\left[E_z^q\left(m+1,n,p+\frac{1}{2}\right)-E_z^q(m,n,p+\frac{1}{2})\right]-\frac{\Delta t}{\mu\Delta z}\left[E_x^q\left(m+\frac{1}{2},n,p+1\right)-E_x^q\left(m+\frac{1}{2},n,p\right)\right]\right\} \quad (11)$$

$$H_z^{q+\frac{1}{2}}\left(m+\frac{1}{2},n+\frac{1}{2},p\right)=H_z^{q-\frac{1}{2}}\left(m+\frac{1}{2},n+\frac{1}{2},p\right)+\left\{\frac{\Delta t}{\mu\Delta y}\left[E_x^q\left(m+\frac{1}{2},n+1,p\right)-E_x^q\left(m+\frac{1}{2},n,p\right)\right]-\frac{\Delta t}{\mu\Delta x}\left[E_y^q\left(m+1,n+\frac{1}{2},p\right)-E_y^q\left(m,n+\frac{1}{2},p\right)\right]\right\} \quad (12)$$

$$E_x^{q+1}\left(m+\frac{1}{2},n,p\right)=\frac{1-\frac{\sigma\Delta t}{2\varepsilon}}{1+\frac{\sigma\Delta t}{2\varepsilon}}E_x^q\left(m+\frac{1}{2},n,p\right)+\frac{1}{1+\frac{\sigma\Delta t}{2\varepsilon}}\left\{\frac{\Delta t}{\varepsilon\Delta y}\left[H_z^{q+\frac{1}{2}}\left(m+\frac{1}{2},n+\frac{1}{2},p\right)-H_z^{q+\frac{1}{2}}\left(m+\frac{1}{2},n-\frac{1}{2},p\right)\right]-\frac{\Delta t}{\varepsilon\Delta z}\left[H_y^{q+\frac{1}{2}}\left(m+\frac{1}{2},n,p+\frac{1}{2}\right)-H_y^{q+\frac{1}{2}}\left(m+\frac{1}{2},n,p-\frac{1}{2}\right)\right]\right\} \quad (13)$$

$$E_y^{q+1}\left(m,n+\frac{1}{2},p\right)=\frac{1-\frac{\sigma\Delta t}{2\varepsilon}}{1+\frac{\sigma\Delta t}{2\varepsilon}}E_y^q\left(m,n+\frac{1}{2},p\right)+\frac{1}{1+\frac{\sigma\Delta t}{2\varepsilon}}\left\{\frac{\Delta t}{\varepsilon\Delta z}\left[H_x^{q+\frac{1}{2}}\left(m,n+\frac{1}{2},p+\frac{1}{2}\right)-H_x^{q+\frac{1}{2}}\left(m,n+\frac{1}{2},p-\frac{1}{2}\right)\right]-\frac{\Delta t}{\varepsilon\Delta x}\left[H_z^{q+\frac{1}{2}}\left(m+\frac{1}{2},n+\frac{1}{2},p\right)-H_z^{q+\frac{1}{2}}\left(m-\frac{1}{2},n+\frac{1}{2},p\right)\right]\right\} \quad (14)$$

To achieve precise results in the finite difference simulation, the grid spacing $\delta$ must be less than the wavelength, normally less than $\lambda/10$. The condition for stability is when step size $\delta$ is the same in all directions ($\delta = \Delta x = \Delta y = \Delta z$)" [16].

$$\Delta t \leq \frac{\delta}{\sqrt{n}c_0} \quad (15)$$

### 2.4. SAR calculation

The SAR can be defined (in terms of time derivatives) by the incremental energy ($dW$) that is absorbed by or dissipated in an incremental mass ($dm$), contained in a volume element ($dV$) of a given mass density ($\rho$) [17].

$$SAR = \frac{d}{dt}\left(\frac{dW}{dm}\right) = \frac{d}{dt}\left(\frac{dW}{\rho dV}\right) \quad (16)$$

$$PLD = \frac{d}{dt}\left(\frac{dW}{dV}\right) \quad (17)$$

The SAR value is expressed in units of watts per kilogram (W/kg). The power loss density (PLD) value is expressed in units of watts per cubic meter (W/m$^3$). The total and maximum point SAR in addition to the maximum electric and magnetic fields were calculated. Total SAR is the PLD of a biological material divided by its mass density, where maximum point SAR is the maximum point SAR of all grid cells. For each grid cell, its point SAR is calculated by its absorbed power divided by its mass.

### 3. Results

**3.1. For vertical polarization**

The maximum values of electric (E) field induced in water in different container models for vertical polarization at 300, 900, and 2,400 MHz with respect to azimuth and elevation angles are shown in Figures 3-6 and both E and H fields are summarized in Table 3. The total SAR values for pyramidal, rectangular, cylindrical and square models are summarized in Tables 4. The maximum value of the electric field induced in water in the pyramidal model for vertical polarization was the highest (9.07, 4.53, and 2.76 V/m at 300, 900, and 2,400 MHz, respectively) followed by the cylindrical model (with 3.68, 2.62, and 1.32 V/m at 300, 900, and 2,400 MHz, respectively) as shown in Figures 3-6 and summarized in Table 3. The maximum value of the electric field induced in water in the rectangular model with 2.27 V/m at 300 MHz was higher than the maximum value of the electric field induced in water in the square model 2.18 V/m. The maximum value of the electric field induced in water in the rectangular model at 900 and 2,400

MHz with 2.11 and 1.09 V/m, respectively, was lower than the maximum value of the electric field induced in water in the square model 2.20 and 1.16 V/m at 900 and 2,400 MHz, respectively.

Table 3: Maximum value of E and H fields induced in water in pyramidal, cylindrical, rectangular and square models for vertical polarization at 300, 900 and 2400 MHz with respect to θ and φ angles.

| Model | Frequency (MHz) | Max. E-field (V/m) | Max. H-field (A/m) | E-field θ–φ configuration | H-field θ–φ configuration |
|---|---|---|---|---|---|
| Pyramidal | 300 | 9.07 | 0.00727 | 0° - 40° | 20° - 40° |
| Cylindrical | 300 | 3.68 | 0.00637 | 40° - 180° | 100° - 40° |
| Rectangular | 300 | 2.27 | 0.00837 | 100° - 40° | 100° - 40° |
| Square | 300 | 2.18 | 0.00638 | 0° - 40° | 100° - 40° |
| Pyramidal | 900 | 4.53 | 0.0108 | 40° - 40° | 0° - 40° |
| Cylindrical | 900 | 2.62 | 0.0083 | 0° - 80° | 0° - 20° |
| Rectangular | 900 | 2.11 | 0.0102 | 0° - 40° | 80° - 40° |
| Square | 900 | 2.20 | 0.0101 | 40° - 180° | 100° - 60° |
| Pyramidal | 2400 | 2.76 | 0.01608 | 20° - 40° | 20° - 140° |
| Cylindrical | 2400 | 1.32 | 0.00878 | 40° - 180° | 40° - 100° |
| Rectangular | 2400 | 1.09 | 0.01258 | 0° - 40° | 20° - 140° |
| Square | 2400 | 1.16 | 0.01316 | 40° - 40° | 40° - 140° |

Although the maximum value of the magnetic field induced in water in the rectangular model was the highest—0.00837 A/m at 300 MHz, compared with 0.00727, 0.00638, and 0.00637 A/m in the pyramidal, square, and cylindrical models, respectively—the pyramidal model produced the highest magnetic fields of 0.0108 and 0.01608 A/m at 900 and 2,400 MHz, respectively, followed by the square, rectangular, and cylindrical models as shown in Table 3. The total SAR induced in water in the pyramidal, cylindrical, rectangular, and square models due to the exposure to a plane wave of 1 V/m amplitude at 300, 900, and 2,400 MHz is summarized in Table 4.

The results indicate that the azimuth angles of propagation direction did not show significant variations in the SAR values induced in water in the four container models, so the effect of azimuth angles on SAR values can be neglected during the representation of the results. The changes in the direction of incident propagation from 0° to 100° with step width 20° relative to vertical polarization cause a significant difference in the observed total SAR values in the four water container models, as shown in Table 4. The tilted angles of incident plane waves induced about 112%, 22%, 63%, and 12% difference in the pyramidal, cylindrical, rectangular, and square models at 300 MHz, respectively, as shown in Table 5. It was observed that the percentage increase in total SAR for the pyramidal model was the highest at 112%, followed by the rectangular model at 300 MHz. For 900 MHz, about 34%, 35%, 24%, and 39% difference was induced in the pyramidal, cylindrical, rectangular, and square models, respectively. For 2,400 MHz, about 30%, 21%, 24%, and 39% difference was induced in the pyramidal, cylindrical, rectangular, and square models, respectively. The results showed that the order of the four container models in terms of the effect of elevation angles (θ) from 0° to 100° on the percentage increase of total SAR for vertical polarization is cylindrical < square < rectangular < pyramidal model at 300, 900, and 2,400 MHz.

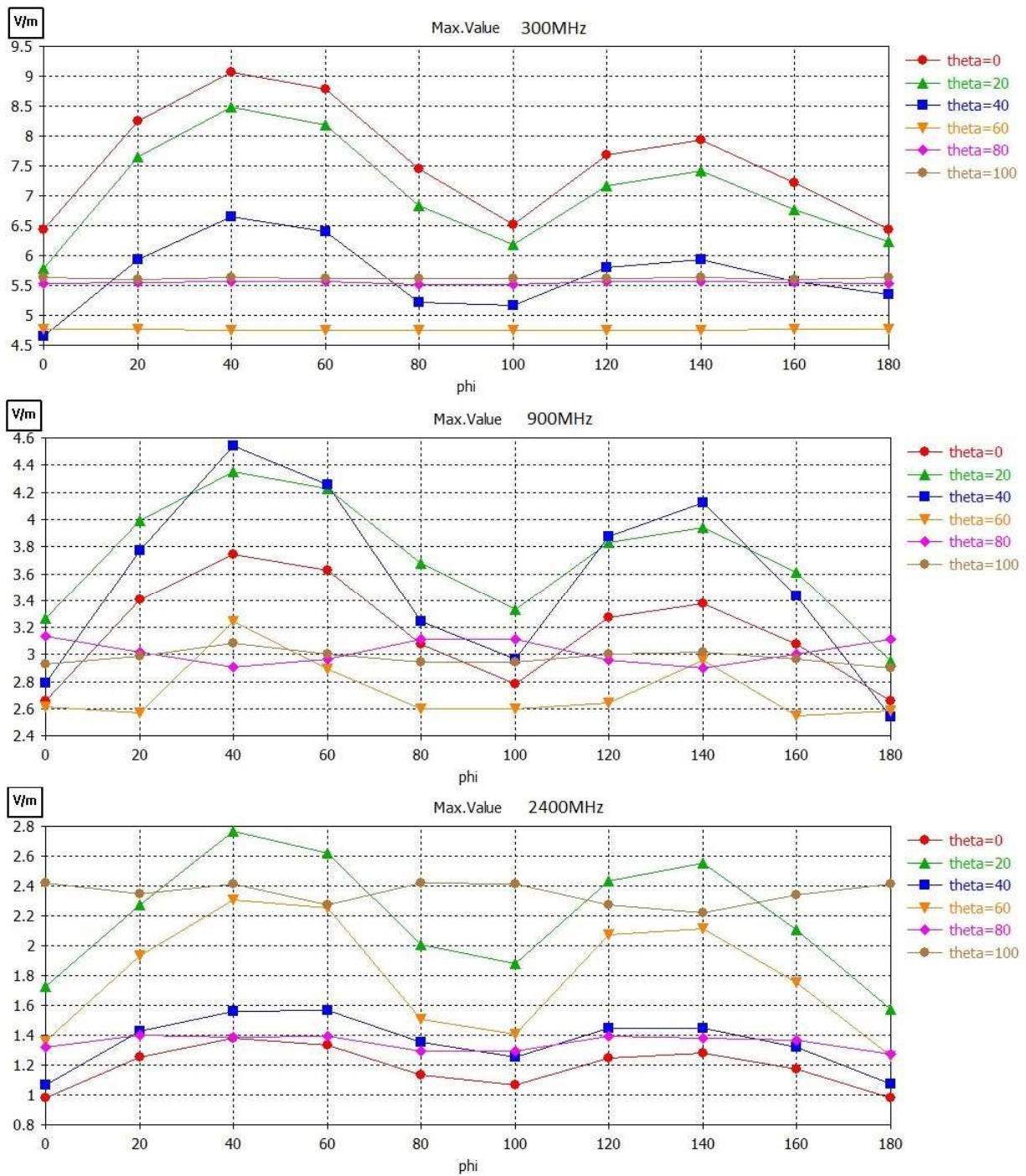

**Figure 3** Maximum value of E field induced in water in pyramidal container model for vertical polarization at 300, 900 and 2400 MHz with respect to θ and φ angles.

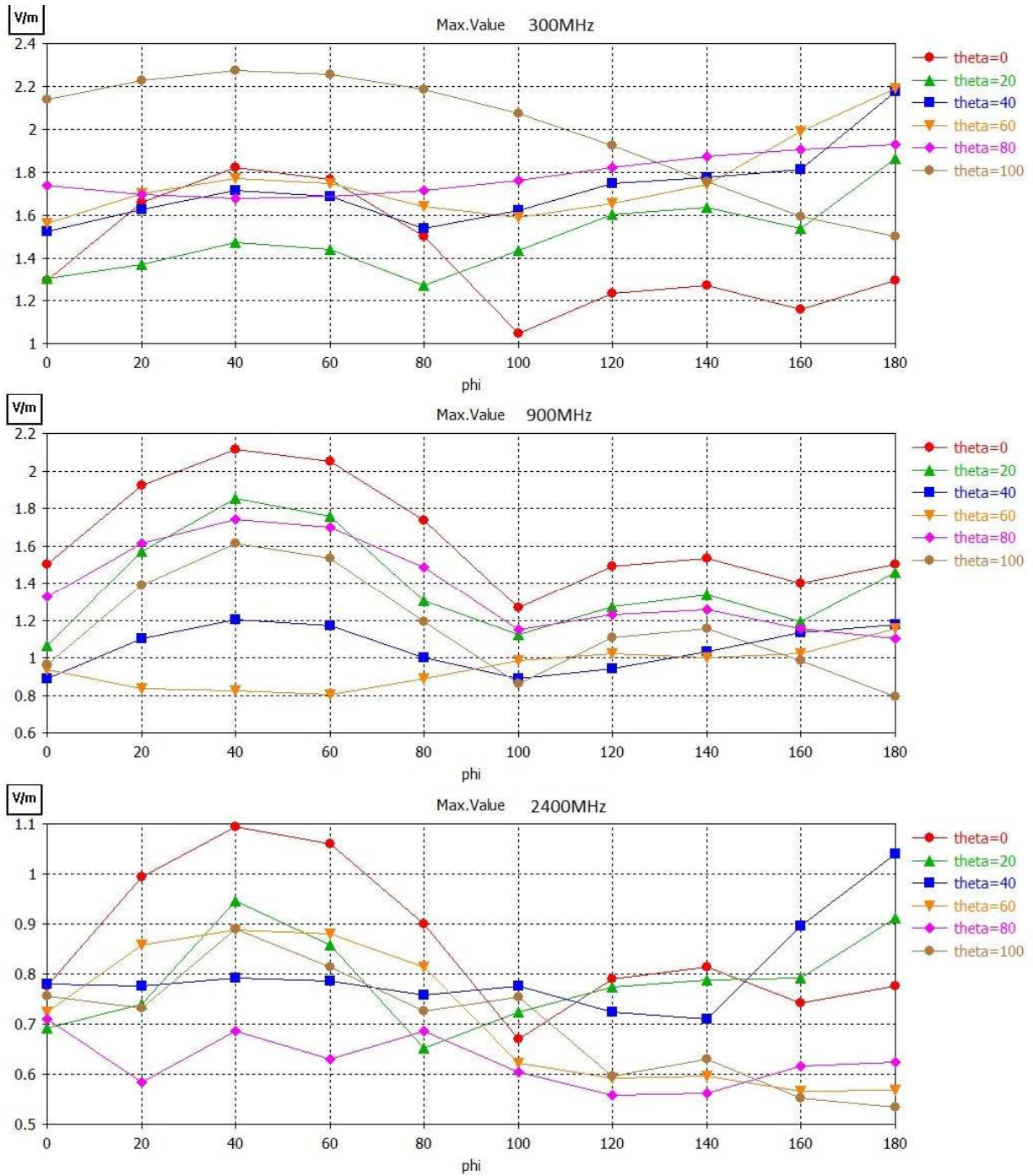

**Figure 4** Maximum value of E field induced in water in rectangular container model for vertical polarization at 300, 900 and 2400 MHz with respect to θ and φ angles.

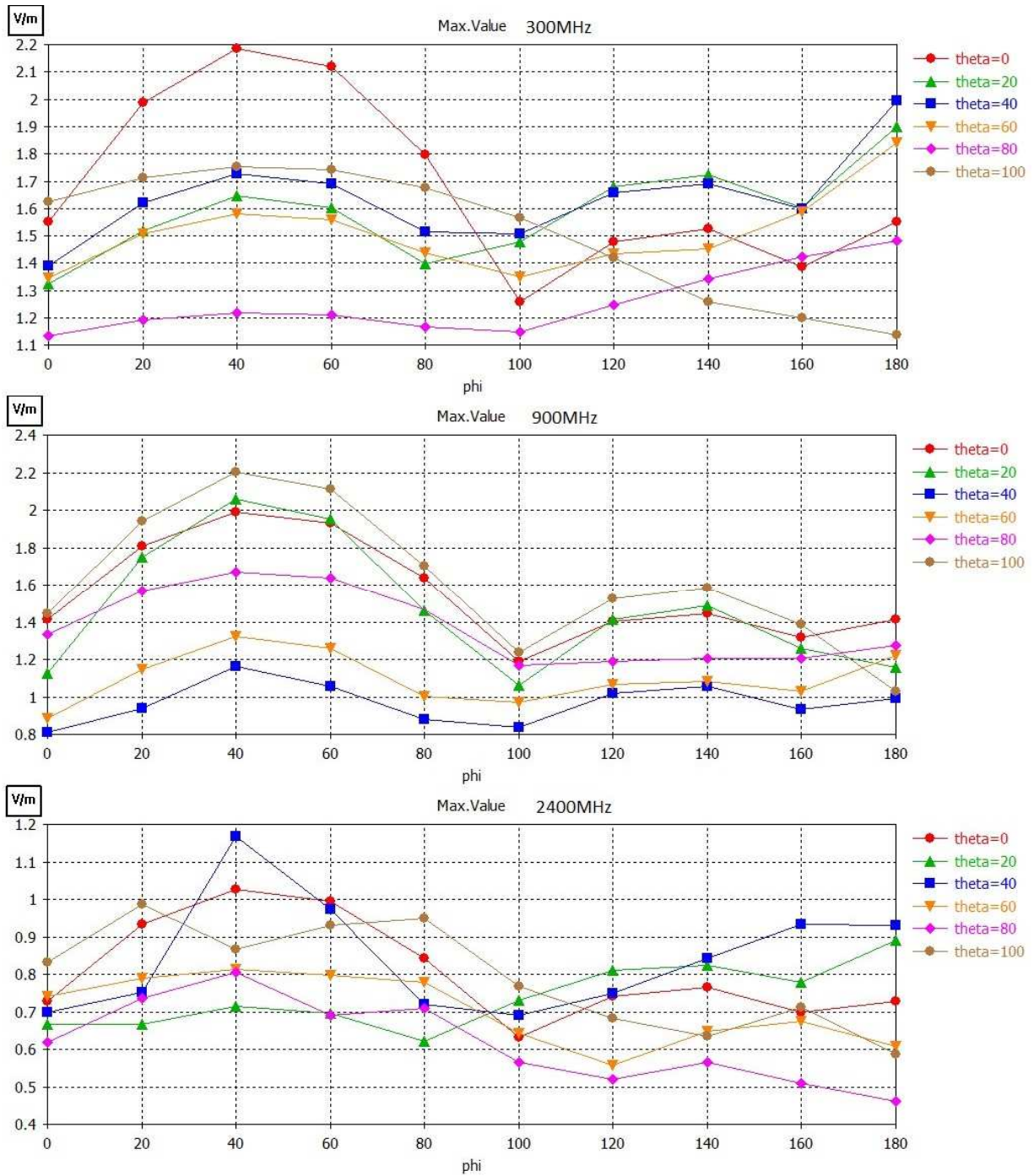

**Figure 5** Maximum value of E field induced in water in square container model for vertical polarization at 300, 900 and 2400 MHz with respect to θ and φ angles.

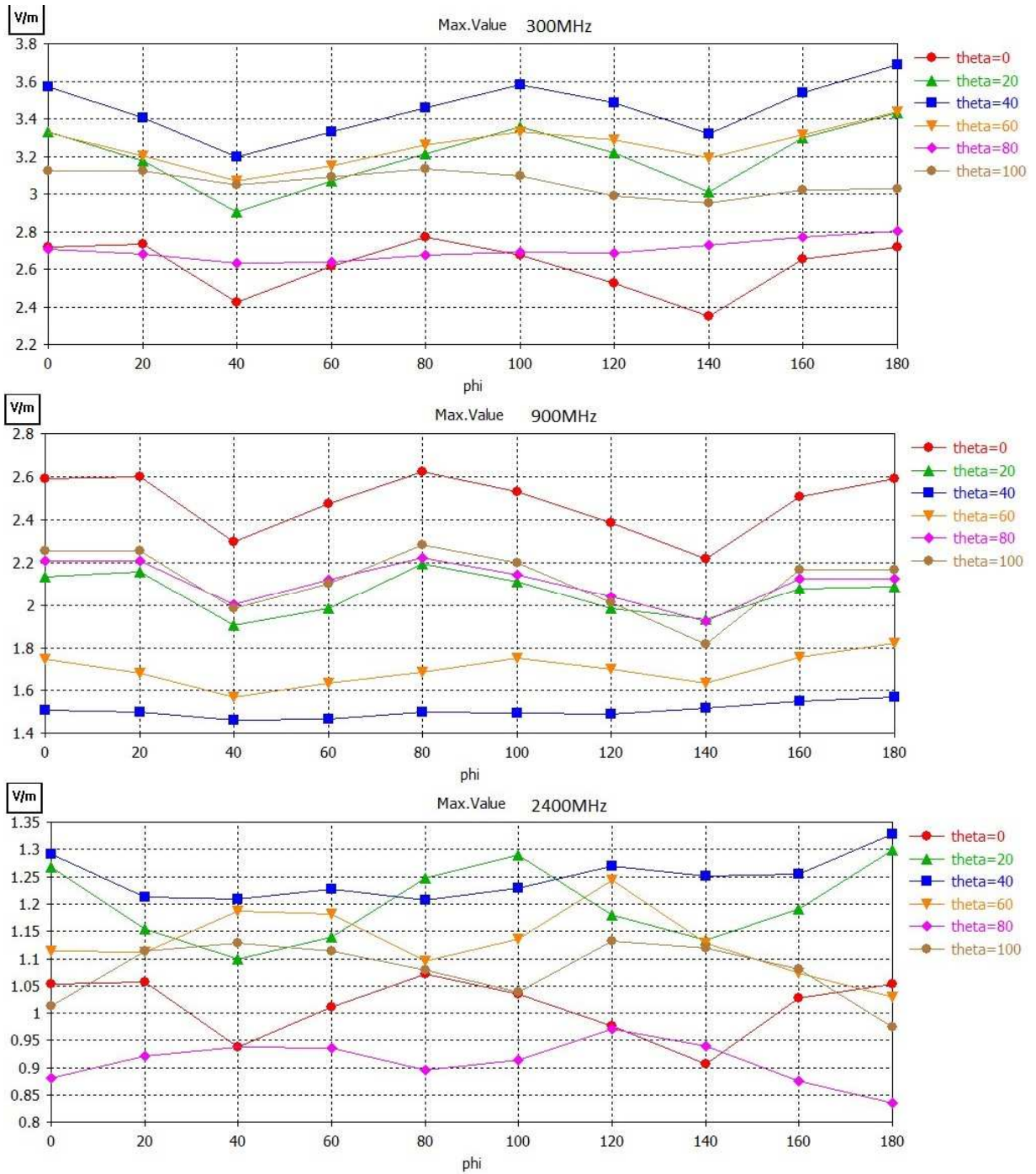

**Figure 6** Maximum value of E field induced in water in cylindrical container model for vertical polarization at 300, 900 and 2400 MHz with respect to θ and φ angles.

Table 4: Maximum values of total SAR induced in water in pyramidal, cylindrical, rectangular and square models for vertical polarization at 300, 900 and 2400 MHz with respect to θ and φ angles.

| Model | Frequency (MHz) | Total SAR (W/ kg) | Theta (θ) | Phi (φ) |
|---|---|---|---|---|
| Pyramidal | 300 | 1.6525e-005 | 0° | 100° |
| Cylindrical | 300 | 9.5251e-006 | 100° | 60° |
| Rectangular | 300 | 1.3977e-005 | 100° | 40° |
| Square | 300 | 9.6561e-006 | 0° | 100° |
| Pyramidal | 900 | 1.599e-005 | 20° | 180° |
| Cylindrical | 900 | 1.417e-005 | 0° | 0° |
| Rectangular | 900 | 1.5366-005 | 0° | 100° |
| Square | 900 | 1.5240e-005 | 0° | 0° |
| Pyramidal | 2400 | 1.3782e-005 | 0° | 0° |
| Cylindrical | 2400 | 1.0401e-005 | 40° | 40° |
| Rectangular | 2400 | 1.2049e-005 | 40° | 180° |
| Square | 2400 | 1.2113e-005 | 40° | 180° |

Table 5: Effect of elevation angles (theta=0°-100°) of the incident plane waves on total SAR at 300, 900 and 2400MHz for vertical polarization:

| Model | Frequency (MHz) | Observed deviation in SAR (mW/Kg) | Percentage increase (%) |
|---|---|---|---|
| Pyramidal | 300 | 0.00778 to 0.0165 | 112 |
| Cylindrical | 300 | 0.00775 to 0.00952 | 22 |
| Rectangular | 300 | 0.00852 to 0.0139 | 63 |
| Square | 300 | 0.00855 to 0.00965 | 12 |
| Pyramidal | 900 | 0.0118 to 0.0159 | 34 |
| Cylindrical | 900 | 0.0104 to 0.0141 | 35 |
| Rectangular | 900 | 0.0123 to 0.0153 | 24 |
| Square | 900 | 0.0110 to 0.0153 | 39 |
| Pyramidal | 2400 | 0.0105 to 0.0137 | 30 |
| Cylindrical | 2400 | 0.00853 to 0.0104 | 21 |
| Rectangular | 2400 | 0.0123 to 0.0153 | 24 |
| Square | 2400 | 0.0110 to 0.0153 | 39 |

The total SAR values induced in water in the pyramidal model showed a gradual decrease by increasing the elevation angle ($\theta$) at 300, 900, and 2,400 MHz. The rectangular model showed a gradual increase in the total SAR values by increasing the elevation angles ($\theta$) at 300 MHz. The highest values of the electric field were obtained at the elevation angles $\theta$ 0°, 40°, and 20° at 300, 900, and 2,400 MHz, respectively. The highest values of magnetic field were recorded at $\theta$ 20°, 0°, and 20° at 300, 900, and 2,400 MHz, respectively. For total SAR, the highest values were recorded at $\theta$ 0°, 20°, and 0° at 300, 900, and 2,400 MHz, respectively. All the highest values belong to the pyramidal container model.

The point SAR distributions in the pyramidal, rectangular, square, and cylindrical models at 300, 900, and 2,400 MHz for vertical polarization in different cross sections are shown in Figures 7-10 and summarized in Table 6. It was shown that the maximum point SAR induced in water in the pyramidal and square models is decreased by increasing the frequency. The highest values of maximum point SAR were obtained at the edges of the four models. The pyramidal model produced the highest value of maximum point SAR at 300, 900, and 2,400 MHz.

Table 6. Maximum point SAR induced in water in pyramidal, cylindrical, rectangular and square models for vertical polarization at 300, 900 and 2400 MHz.

| Model | Frequency (MHz) | Maximum Point SAR (W/ kg) |
|---|---|---|
| Pyramidal | 300 | 0.007009 |
| Cylindrical | 300 | 0.1837e-03 |
| Rectangular | 300 | 0.001576 |
| Square | 300 | 0.9717e-3 |
| Pyramidal | 900 | 0.001672 |
| Cylindrical | 900 | 0.1933e-03 |
| Rectangular | 900 | 0.9962e-03 |
| Square | 900 | 0.6534e-03 |
| Pyramidal | 2400 | 0.171e-03 |
| Cylindrical | 2400 | 0.1612e-03 |
| Rectangular | 2400 | 0.3815e-03 |
| Square | 2400 | 0.309e-03 |

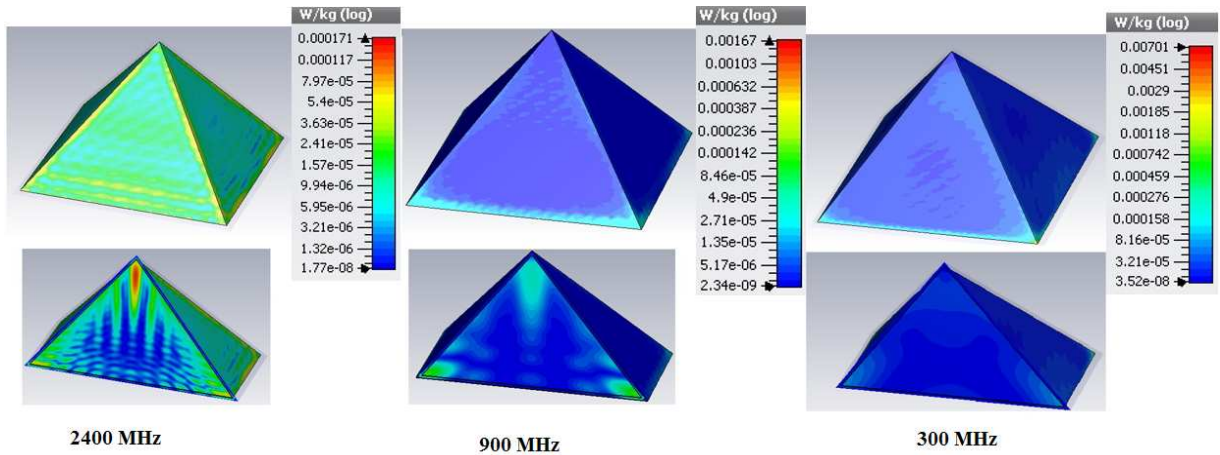

**Figure 7** Point SAR distributions induced in water in pyramidal container model for vertical polarization at 300, 900 and 2400 MHz.

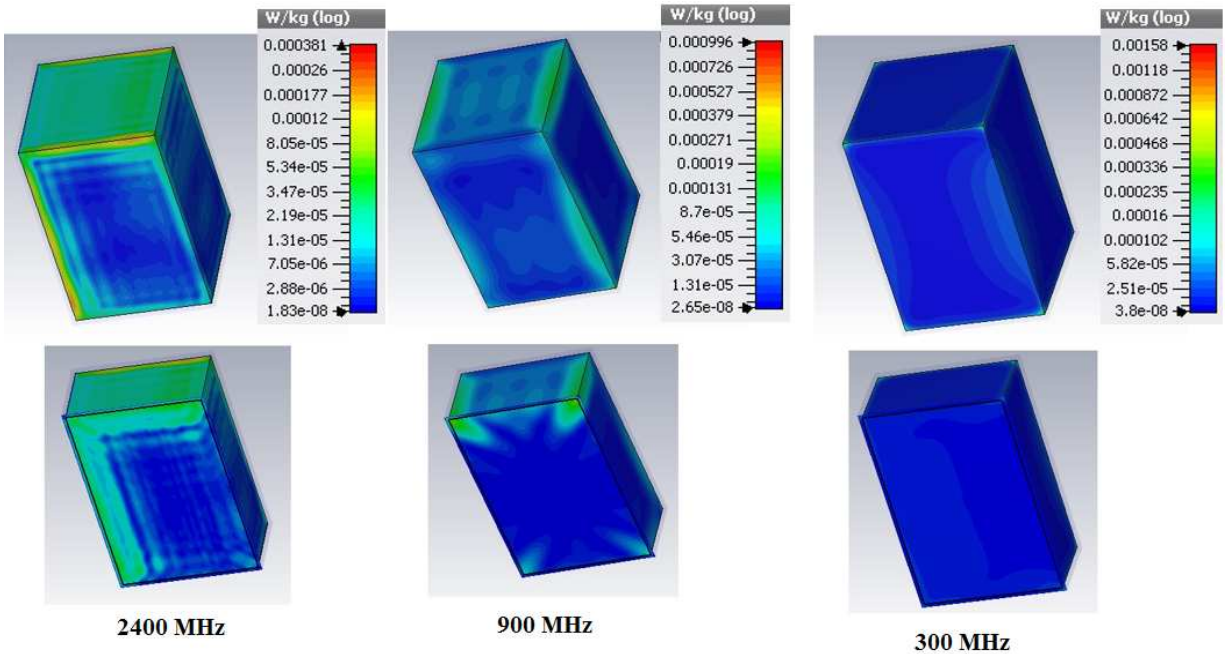

**Figure 8** Point SAR distributions induced in water in rectangular container model for vertical polarization at 300, 900 and 2400 MHz.

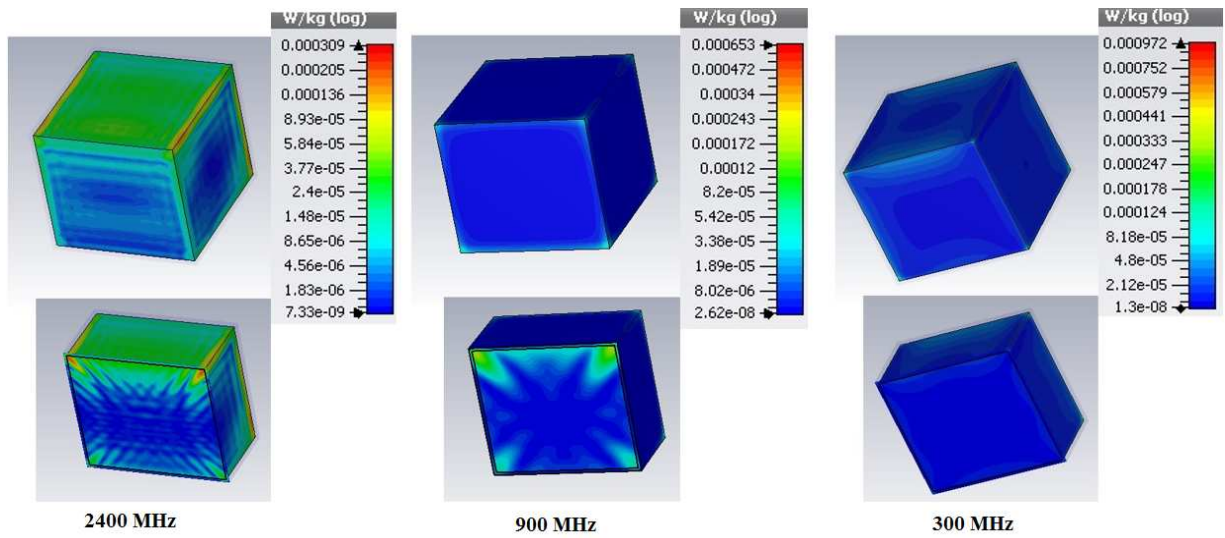

**Figure 9** Point SAR distributions induced in water in square container model for vertical polarization at 300, 900 and 2400 MHz.

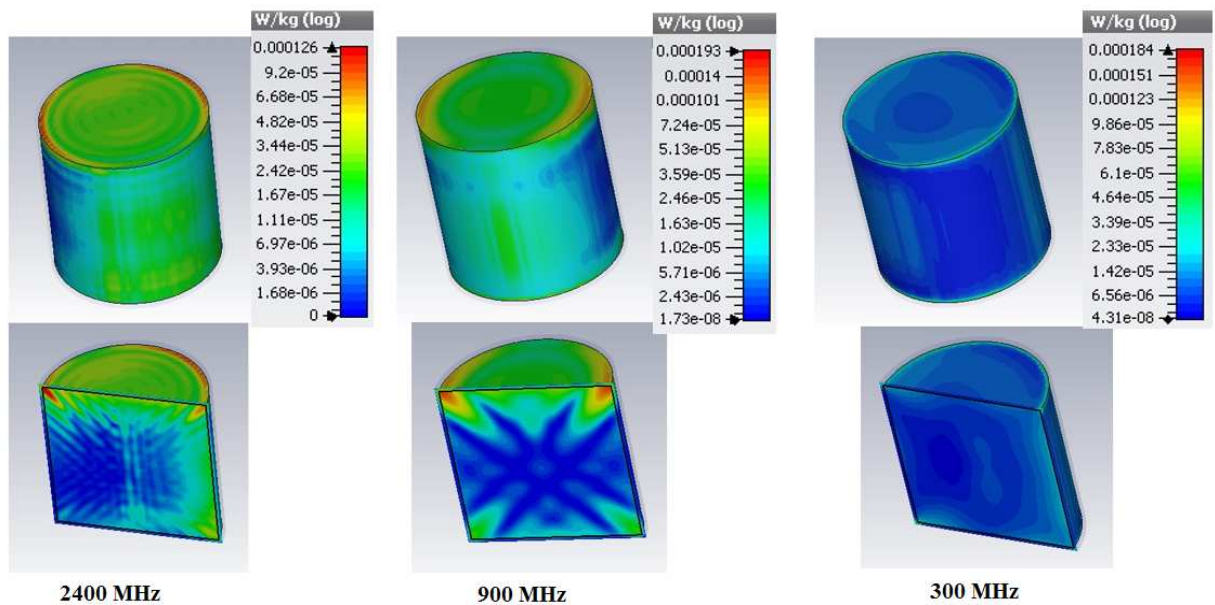

**Figure 10** Point SAR distributions induced in water in cylindrical container model for vertical polarization at 300, 900 and 2400 MHz.

## 3.2. For horizontal polarization

The maximum values of electric (E) field induced in water in different container models for horizontal polarization at 300, 900, and 2,400 MHz with respect to θ and φ angles are shown in Figures 11-14 and both E and H fields are summarized in Table 7 and the total SAR values for pyramidal, rectangular, cylindrical and square models are summarized in Tables 8. Following the same trend, the maximum electric field induced in water in the pyramidal model for horizontal polarization was also the highest with 9.13, 4.13, and 2.34 V/m at 300, 900, and 2,400 MHz, respectively, followed by the cylindrical model with 2.73, 2.62, and 1.42 V/m at 300, 900, and 2,400 MHz, respectively as shown in Table 7. The maximum electric field induced in water was 1.82, 2.15, and 1.15 V/m for the rectangular model and 2.17, 2.12, and 1.20 V/m for the square model at 300, 900, and 2,400 MHz, respectively.

The maximum value of the magnetic field induced in water in the pyramidal model was the highest with 0.0095, 0.012, and 0.0149 A/m at 300, 900, and 2,400 MHz, respectively, followed by the square model with 0.00693, 0.0104, and 0.01319 A/m at 300, 900, and 2,400 MHz, respectively as shown in Table 7. The maximum magnetic field induced in the rectangular model was 0.00612, 0.0101, and 0.01183 A/m at 300, 900, and 2,400 MHz, respectively.

Table 7: Maximum value of E and H fields induced in water in pyramidal, cylindrical, rectangular and square models for horizontal polarization at 300, 900 and 2400 MHz with respect to θ and φ angles.

| Model | Frequency (MHz) | Max. E-field (V/m) | Max. H-field (A/m) | E-field θ–φ configuration | H-field θ–φ configuration |
|---|---|---|---|---|---|
| Pyramidal | 300 | 9.13 | 0.0095 | 0° - 140° | 100° - 0° |
| Cylindrical | 300 | 2.73 | 0.0056 | 0° - 100° | 40° - 0° |
| Rectangular | 300 | 1.82 | 0.00612 | 0° - 140° | 40° - 0° |
| Square | 300 | 2.17 | 0.00693 | 0° - 140° | 40° - 0° |
| Pyramidal | 900 | 4.13 | 0.012 | 60° - 140° | 80° - 60° |
| Cylindrical | 900 | 2.62 | 0.0087 | 0° - 100° | 20° - 100° |
| Rectangular | 900 | 2.15 | 0.0101 | 20° - 120° | 20° - 0° |
| Square | 900 | 2.12 | 0.0104 | 40° - 100° | 60° - 0° |
| Pyramidal | 2400 | 2.34 | 0.0149 | 80° - 100° | 20° - 160° |
| Cylindrical | 2400 | 1.42 | 0.00947 | 60° - 80° | 40° - 0° |
| Rectangular | 2400 | 1.15 | 0.01183 | 60° - 120° | 100° - 140° |
| Square | 2400 | 1.20 | 0.01319 | 40° - 40° | 40° - 140° |

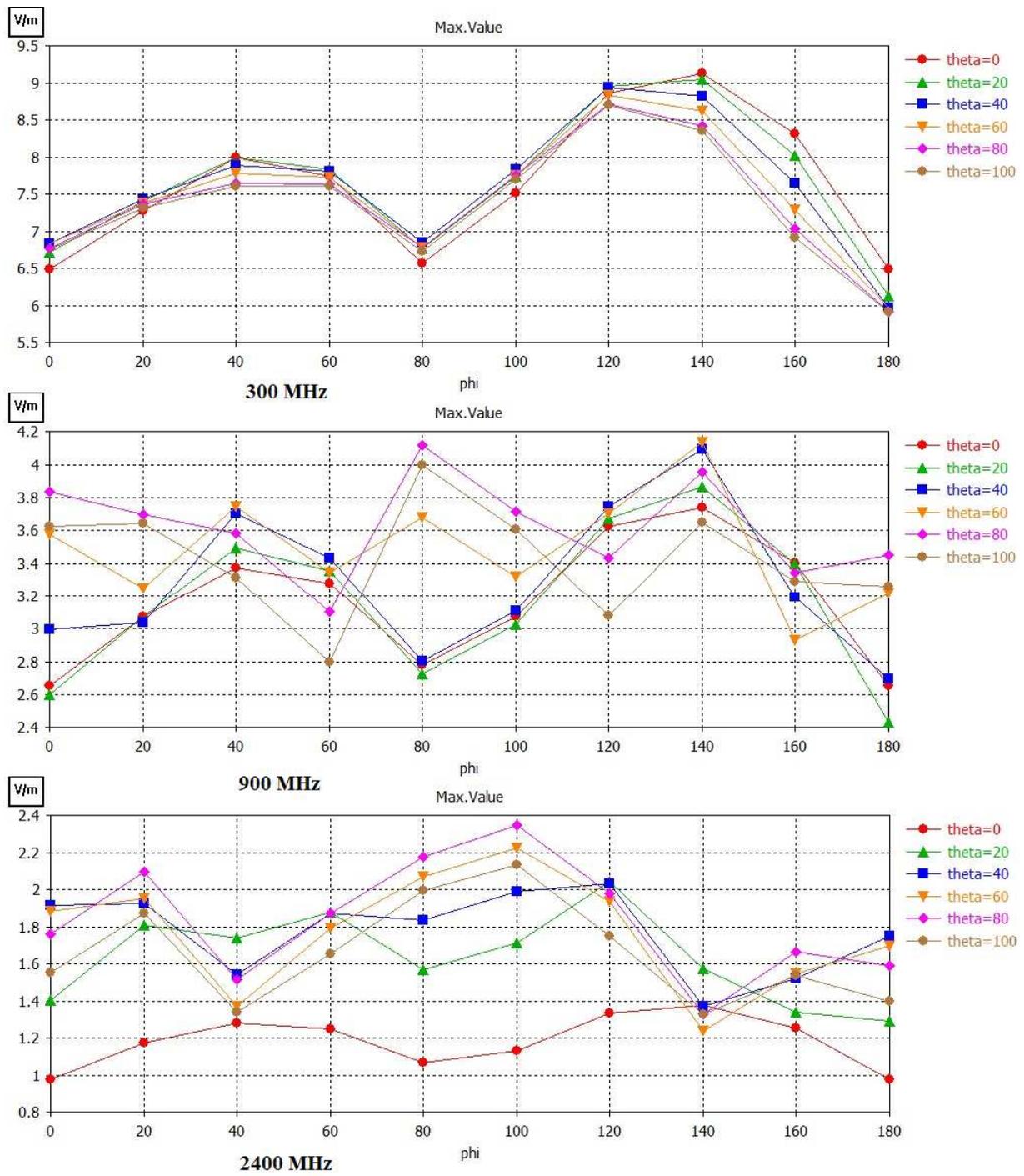

**Figure 11** Maximum value of E field induced in water in pyramidal container model for horizontal polarization at 300, 900 and 2400 MHz with respect to θ and φ angles.

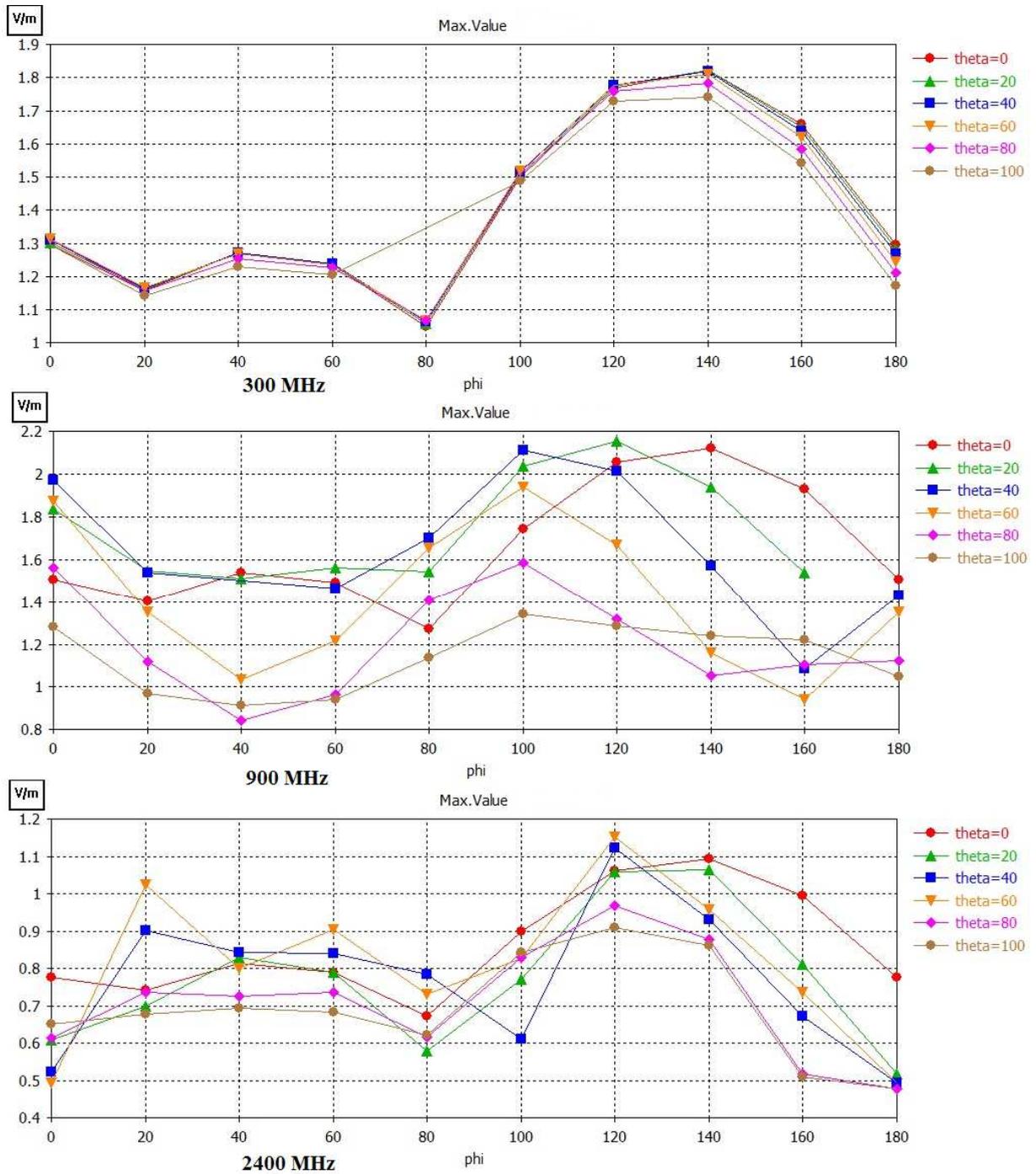

**Figure 12** Maximum value of E field induced in water in rectangular container model for horizontal polarization at 300, 900 and 2400 MHz with respect to θ and φ angles.

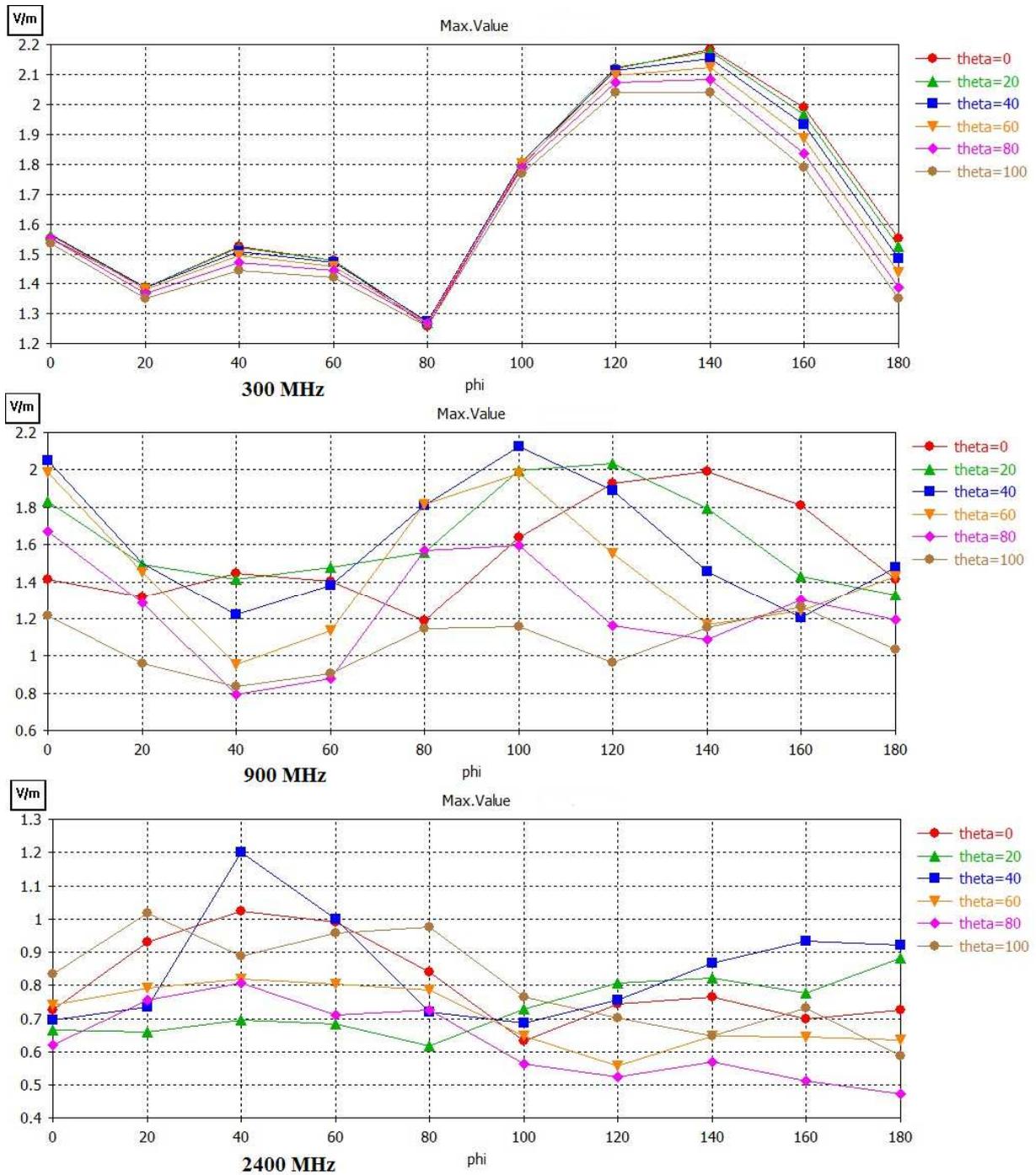

**Figure 13** Maximum value of E field induced in water in square container model for horizontal polarization at 300, 900 and 2400 MHz with respect to θ and φ angles.

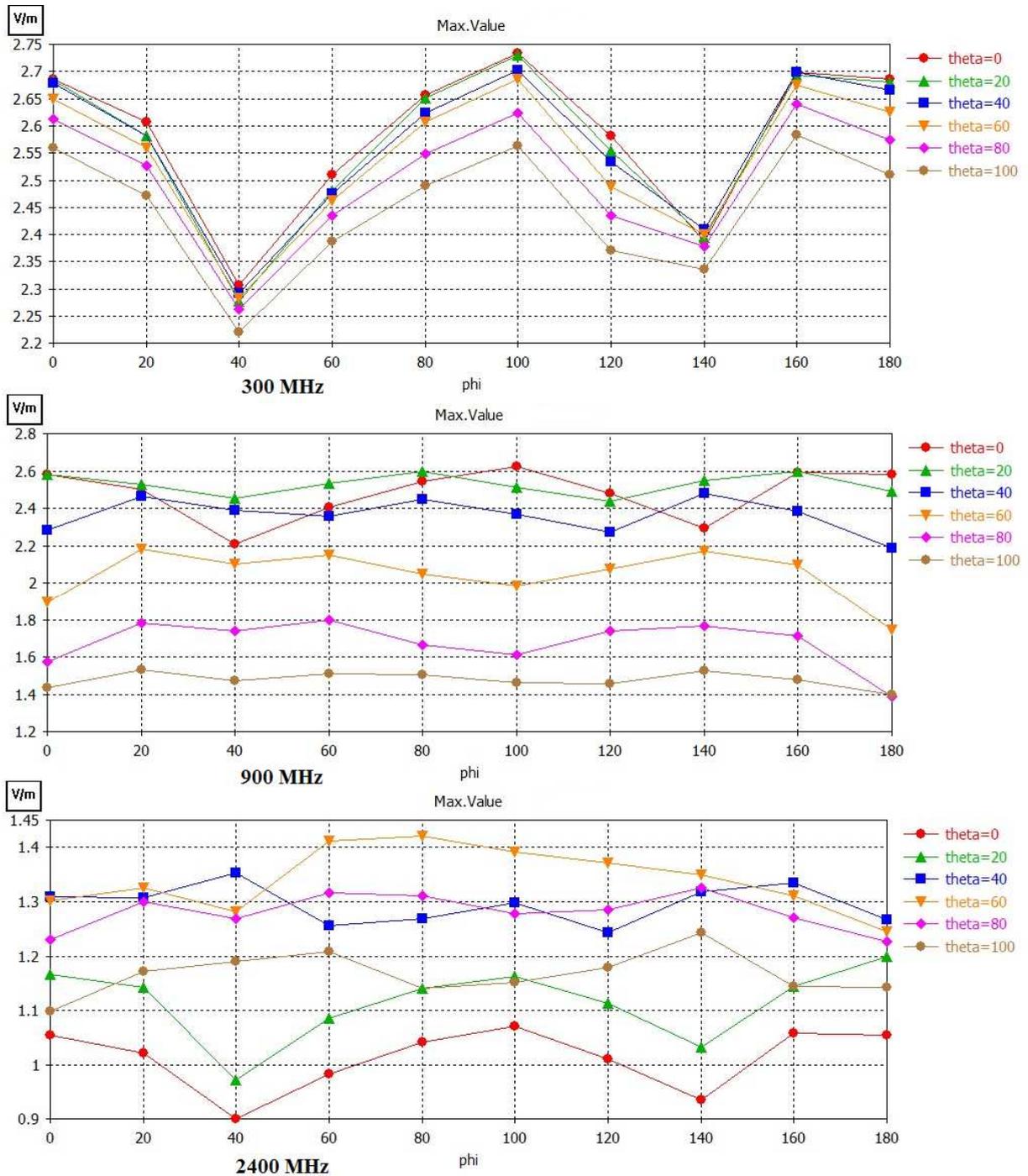

**Figure 14** Maximum value of E field induced in water in cylindrical container model for horizontal polarization at 300, 900 and 2400 MHz with respect to θ and φ angles.

According to the results, the azimuth angles of propagation direction caused sharp deviations in the total SAR values induced in water in the pyramidal and square models at 900 MHz. It was observed that when the elevation ($\theta$) angle of plane waves was increased, the total SAR values in the cylindrical and rectangular models at 300 and 900 MHz showed a gradual declining trend, although the increase of the elevation angle also caused a gradual decrease in the total SAR values in the pyramidal and square models at 2,400 and 900 MHz, respectively. The total SAR values showed an increase in the pyramidal model at 300 and 900 MHz.

Table 8: Maximum value of total SAR induced in water in pyramidal, cylindrical, rectangular and square models for horizontal polarization at 300, 900 and 2400 MHz with respect to θ and φ angles.

| Model | Frequency (MHz) | Total SAR (W/ kg) | Theta (θ) | Phi (φ) |
|---|---|---|---|---|
| Pyramidal | 300 | 2.0873e-005 | 100° | 40° |
| Cylindrical | 300 | 7.7378e-006 | 0° | 0° |
| Rectangular | 300 | 8.5232e-006 | 0° | 80° |
| Square | 300 | 9.886e-006 | 100° | 40° |
| Pyramidal | 900 | 1.854e-005 | 60° | 140° |
| Cylindrical | 900 | 1.409e-005 | 0° | 0° |
| Rectangular | 900 | 1.525e-005 | 0° | 80° |
| Square | 900 | 1.555e-005 | 100° | 0° |
| Pyramidal | 2400 | 1.407e-005 | 20° | 140° |
| Cylindrical | 2400 | 9.860 e-006 | 100° | 180° |
| Rectangular | 2400 | 1.2318e-005 | 100° | 40° |
| Square | 2400 | 1.201e-005 | 40° | 180° |

There are significant differences in the observed total SAR values in the four water container models due to the changes in the elevation angles ($\theta$) from 0° to 100° with a step width of 20° relative to horizontal polarization, as shown in Table 9. The order of the effect of the elevation angles of incident plane waves on total SAR was about 25%, 10%, 3%, and 2% difference in water in the pyramidal, rectangular, square, and cylindrical models at 300 MHz, respectively. For 900 MHz, about 36%, 33%, 31%, and 18% difference was induced in water in the square, rectangular, cylindrical, and pyramidal models, respectively. For 2,400 MHz, about 48%, 39%, 29%, and 21% difference was induced in the pyramidal, rectangular, square, and cylindrical models, respectively. Although the percentage increase in total SAR for the pyramidal and rectangular models were 112% and 63%, respectively, at 300 MHz for vertical polarization, the percentage increase in total SAR for the pyramidal model was the highest, with 48%, followed by the rectangular model, with 39%, at 2,400 MHz for horizontal polarization. The results showed that the order of the container models in terms of the effect of elevation angles ($\theta$) from 0° to 100° on total SAR is cylindrical < square < rectangular < pyramidal model at 300, 900, and 2,400 MHz for both vertical and horizontal polarizations.

Table 9: Effect of elevation angles (theta=0°-100°) of the incident plane waves on total SAR at 300, 900 and 2400MHz for horizontal polarization:

| Model | Frequency (MHz) | Observed deviation in SAR (mW/Kg) | Percentage increase (%) |
|---|---|---|---|
| Pyramidal | 300 | 0.01668 to 0.02087 | 25 |
| Cylindrical | 300 | 0.007598 to 0.0077378 | 2 |
| Rectangular | 300 | 0.007689 to 0.0085232 | 10 |
| Square | 300 | 0.009528 to 0.009886 | 3 |
| Pyramidal | 900 | 0.01568 to 0.01854 | 18 |
| Cylindrical | 900 | 0.01074 to 0.01409 | 31 |
| Rectangular | 900 | 0.01141 to 0.01525 | 33 |
| Square | 900 | 0.0114 to 0.01555 | 36 |
| Pyramidal | 2400 | 0.009497 to 0.01407 | 48 |
| Cylindrical | 2400 | 0.008539 to 0.00986 | 21 |
| Rectangular | 2400 | 0.00884 to 0.012318 | 39 |
| Square | 2400 | 0.009261 to 0.01201 | 29 |

As shown in Figures 15-18, the maximum point SAR distributions in the four container models at 300, 900, and 2,400 MHz for horizontal polarization in different cross sections, and these are summarized in Table 10. The maximum point SAR induced in water in the pyramidal and square models is decreased by increasing the frequency. The highest values of maximum point SAR were also obtained near the edges of the four models. The highest value of maximum point SAR at 300, 900, and 2,400 MHz was induced in the pyramidal model.

Table 10. Maximum point SAR induced in water in pyramidal, cylindrical, rectangular and square models for horizontal polarization at 300, 900 and 2400 MHz.

| Model | Frequency (MHz) | Maximum Point SAR (W/ kg) |
|---|---|---|
| Pyramidal | 300 | 0.009334 |
| Cylindrical | 300 | 0.1428e-03 |
| Rectangular | 300 | 0.6845e-03 |
| Square | 300 | 0.001347 |
| Pyramidal | 900 | 0.002694 |
| Cylindrical | 900 | 0.1932e-03 |
| Rectangular | 900 | 0.9962e-03 |
| Square | 900 | 0.9144e-03 |
| Pyramidal | 2400 | 0.5559e-03 |
| Cylindrical | 2400 | 0.1459e-03 |
| Rectangular | 2400 | 0.4175e-03 |
| Square | 2400 | 0.4289e-03 |

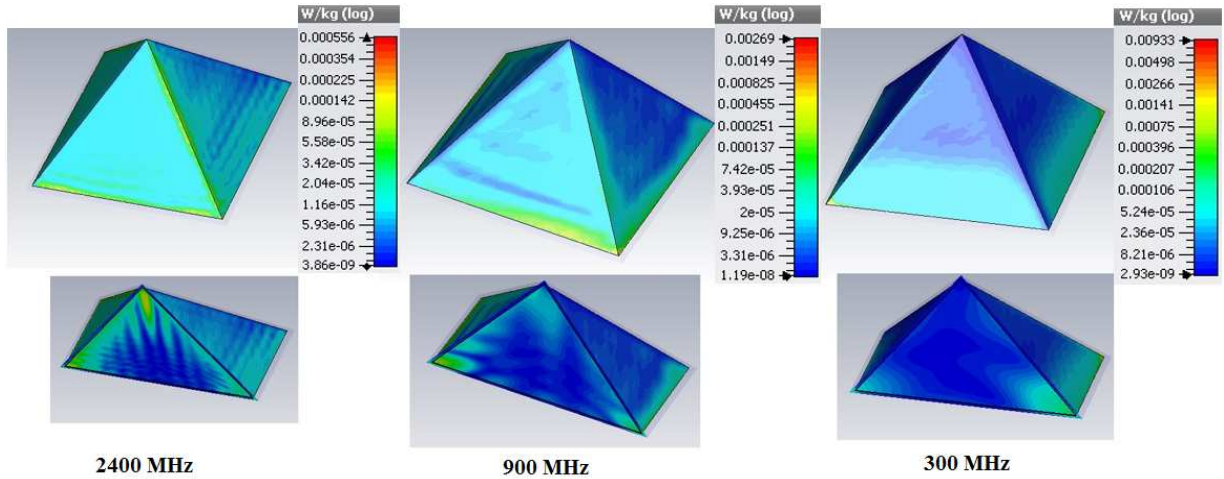

**Figure 15** Point SAR distributions induced in water in pyramidal container model for horizontal polarization at 300, 900 and 2400 MHz.

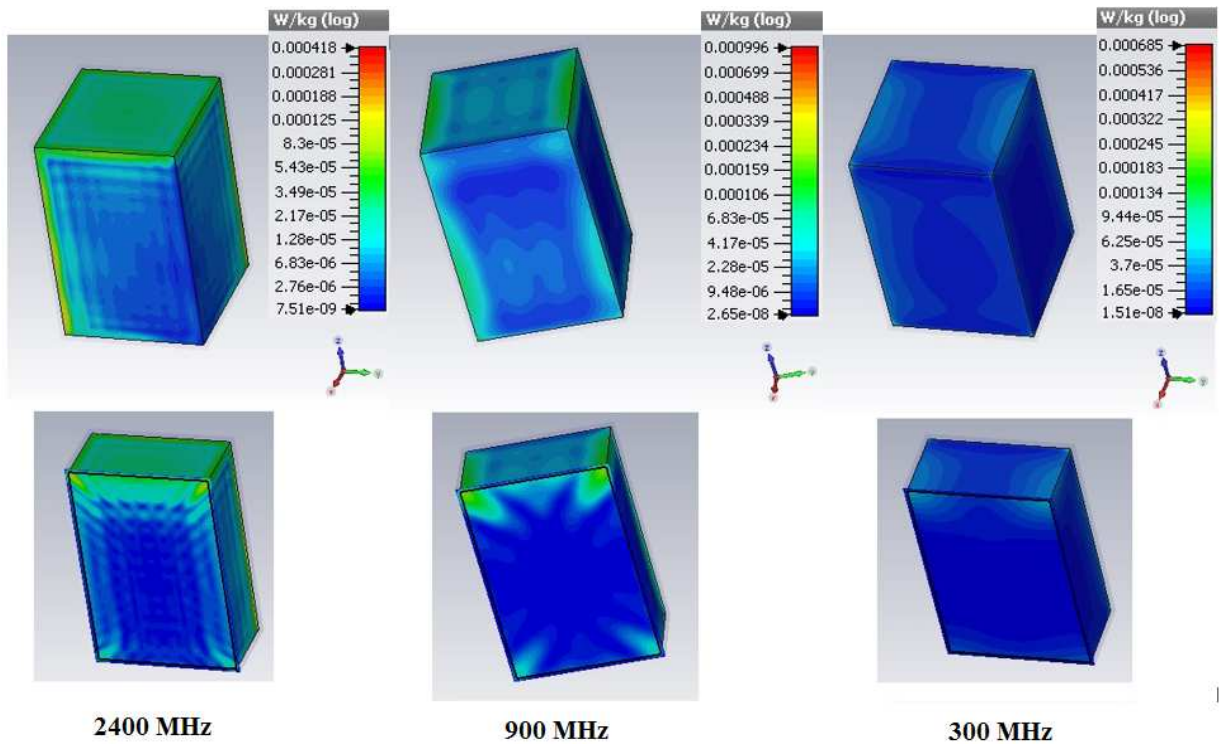

**Figure 16** Point SAR distributions induced in water in rectangular container model for horizontal polarization at 300, 900 and 2400 MHz.

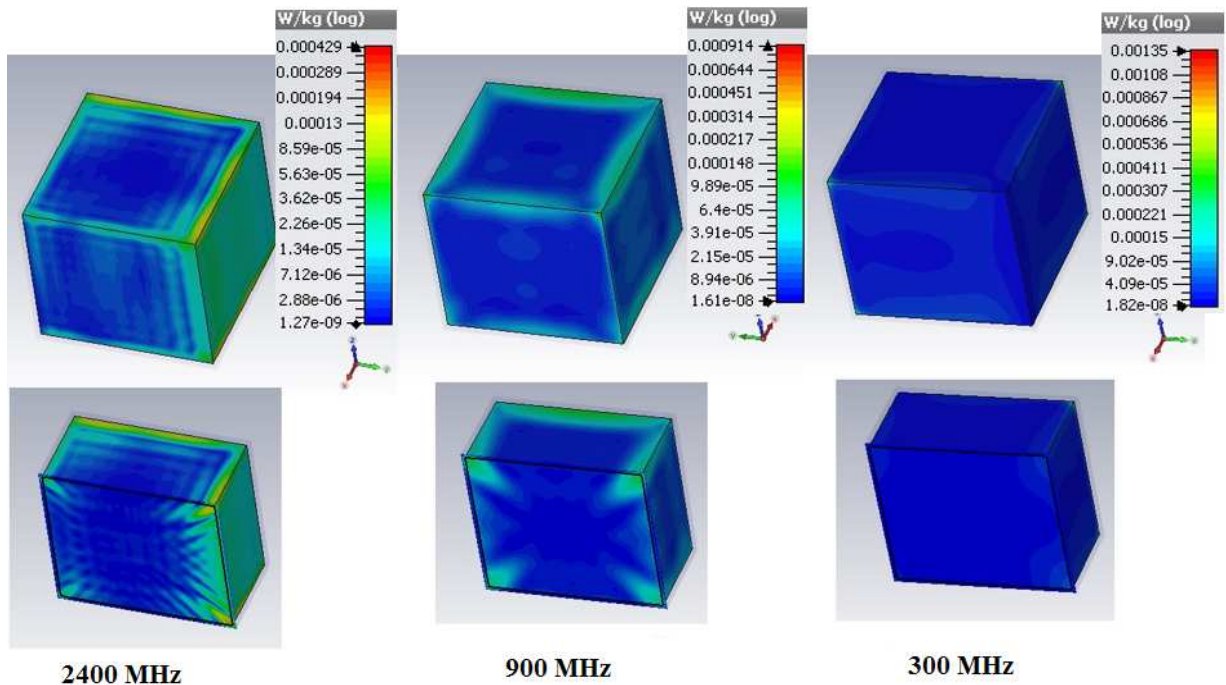

**Figure 17** Point SAR distributions induced in water in square container model for horizontal polarization at 300, 900 and 2400 MHz.

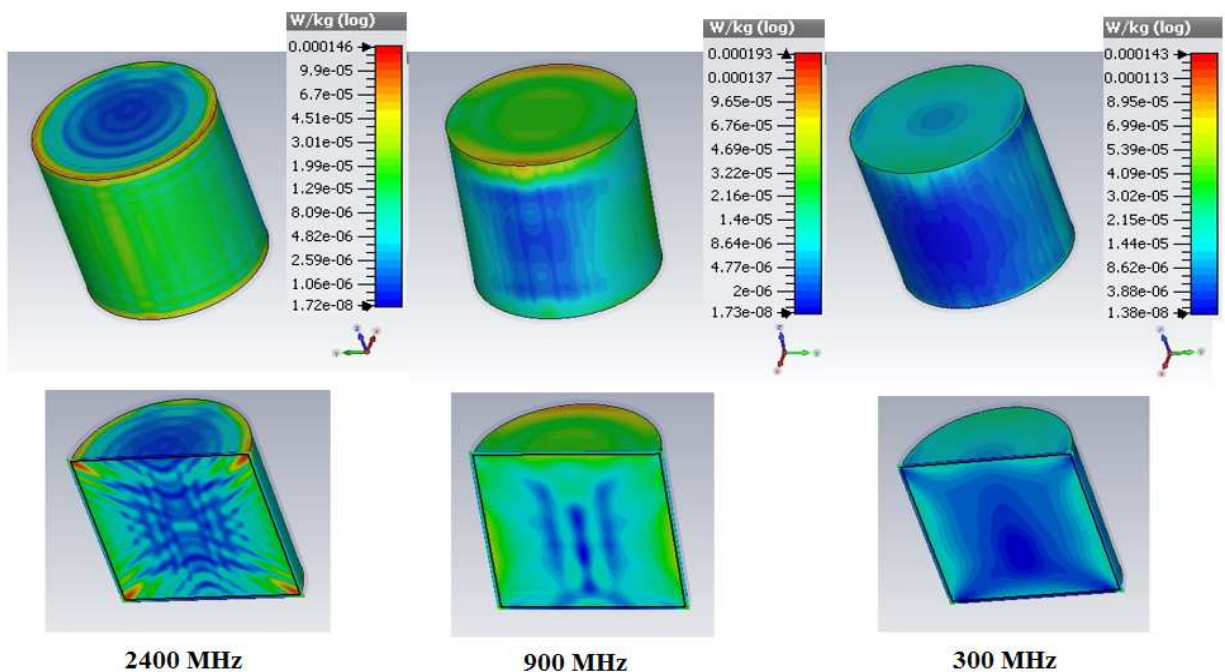

**Figure 18** Point SAR distributions induced in water in cylindrical container model for horizontal polarization at 300, 900 and 2400 MHz.

According to the results of the maximum electric and magnetic fields, total SAR, and maximum point SAR, There are some points that require close attention, have been obtained:

- The highest values of electric and magnetic fields, total SAR, and maximum point SAR induced in water were observed in the pyramidal model.
- The highest values of electric and magnetic fields and total SAR for 300, 900, and 2,400 MHz for vertical polarization were observed at the elevation angles range 0°–40°.
- The azimuth angles range from 0° to 180° does not cause a significant effect on the total SAR values for vertical polarization.
- For vertical polarization, the highest values of electric and magnetic fields and total SAR were obtained for 300, 900, and 2,400 MHz at the θ–φ configuration 0°–140°, 60°–140°, 80°–100°; 100°–0°, 80°–60°, 20°–160°; and 100°–40°, 60°–140°, 20°–140°, respectively.
- The azimuth angles range from 0° to 180° caused a significant change in the total SAR values for horizontal polarization in the pyramidal model at 900 MHz, the rectangular model at 900 and 2,400 MHz, and the square model at 900 and 2,400 MHz.
- The maximum point SAR induced in water in the pyramidal model for horizontal polarization was higher than for the vertical polarization.

After studying the behavior of EM waves at 300, 900, and 2400 MHz—which is environmentally abundant relative to vertical and horizontal polarizations over a wide range of elevation angles of 0°–100° and azimuth angles range of 0°–180° for vertical and horizontal polarizations and its relationship to the packaging shape of water containers, it was shown that the variation in the packaging shape of water containers induced different electric and magnetic fields, total SAR and maximum point SAR values in the stored water.

## 4. Discussion

**4.1 Mechanisms of effect**

The relationship and the interaction between EM waves in the microwave frequency range and water solutions were explained by the well-known mechanisms of ionic conduction and dipole rotation [18-20]. The movement of charged particles dissolved in water towards the opposite charged plate in the ionic conduction process explains how a high-frequency change in the polarity of the electric field of EM radiation causes friction and collision between the charged particles and water molecules, which could affect the strength of the hydrogen bonding network [20; 21]. In dipole rotation mechanism, the oscillation of the electric field component of EM waves in the microwave frequency range is more than one billion cycles per second, and because of the dipole molecule of water, the molecules try to align with the applied field, which in turn affects the strength of the hydrogen bonding network and generates heat when water is exposed to EM radiation with high power density [19; 20].

Recent studies showed that there is a change in the viscosity and conductivity of H2O-NaCl solution because of the effect of EM radiation on hydrogen bonds caused by the dipole rotation mechanism after the exposure to nonthermal low-intensity EM radiation at microwave frequencies of 450 and 2,400 MHz [10; 11]. The study [10] considered the effect of EM radiation on diffusion in H2O-NaCl solution as a possible mechanism for the nonthermal effects. The nonthermal effects of EM fields were also confirmed by more than 1,800 new studies reported in the BioInitiative 2012 Report [22]. For example, SAR 0.00015–0.003 W/kg affects calcium ion movement [23], SAR 0.000021–0.0021 W/kg affects cell proliferation [24], and SAR 0.0024–0.024 W/kg affects DNA damage [25]. These values of SAR reported in the studies [22], [23], [24] and [25] are in the range of the maximum point SAR values reported in this EM simulation study as shown in tables 6 and 10 for vertical and horizontal polarisations, respectively.

**4.2 Hypothesized mechanism of effect**

The EM wave consists of many small units or segments. Each segment has the same wavelength as the EM wave and is called a photon, which has a dual nature—wave and particle. Photon energy is limited by its frequency, where the photon at a specific frequency has a specific amount of energy. The energy of the EM wave can be expressed by joules and electron volts [26]. The photon behaves as a wave until it strikes an atom or electron and its energy is absorbed and ends up in one point. To represent the worst exposure scenario, the highest values of maximum point SAR obtained from the EM simulation for vertical and horizontal polarisations can be used to calculate the number of photons or the amount of energy at a point in water that absorbed the highest amount of energy. For example, the highest value of maximum point SAR for vertical polarisation as shown in table 6 was 0.007009 W/kg at 300 MHz. This value was obtained during the excitation duration of 1.18484958 nanoseconds using CST STUDIO SUITE 2014 EM simulation software.

The amount of energy absorbed by water at that point SAR during 1.18484958 nanoseconds can be calculated by converting the power absorbed at the highest point SAR to electron volts [27], where watt is equal to joule/second [28; 29].

$$Watt = J/s$$

and

$$1J = 6.241509\mathrm{e}18 \text{ eV}$$

Excitation duration $= 1.18484958e - 9\ s$

Therefore,

0.007009 W/kg $= 0.007009$ (joule / second) * kg $\times\ 1.18484958e - 9\ s$

$$= 8.30461070622e - 12\ (Joule/s) * kg$$

8.30461070622e-12 (Joule/ s)*kg $\times$ *6.241509e18*

$$= 5.19\mathrm{e}7\ eV * \text{kg}$$

The highest value of energy absorbed by water at a point during the excitation duration of 1.18484958 nanoseconds is about *5.19e7 eV* * kg

The energy absorbed at a point in water per kg per nanosecond can be calculated as follows

*5.19e7 eV* * kg /*1.18484958*

$$= 4.38\mathrm{e}7\ eV * \text{kg}$$

The energy at point of water per kg per picosecond is:

$$= 4.38\mathrm{e}7\ \text{eV} * \text{kg} / 1000$$

$$= 43800\ \text{eV} * kg$$

It has been reported that the energy required to break the hydrogen bonding between water molecules, which is called bond-dissociation energy (BDE), is about 23 kJ/mol, and it is equal to 0.238 eV [30]. The hydrogen bonding lifetime ranges between 1 ps and 20 ps [31].

The calculation above shows that the highest amount of EM energy absorbed at a point of water obtained from EM simulation is *43800 eV* * kg, which is 184,000 times higher than the energy required to break the hydrogen bonds. This amount of energy can theoretically explain how EM waves at 300 MHz are capable of breaking the hydrogen bonding of water if the *43800 eV \* kg* is distributed over 184,000 water molecules or hydrogen bonds, but the amount of energy is expressed in *eV \* kg*.

It is well known that electric field intensity is used to determine the location where the photons strike the electrons or atoms. The photons strike where the electric field intensity is highest (electric field strength)$^2$ ($E_{max}$)$^2$. The location of striking is proportional to the field intensity at that location. According to the EM simulation results of point SAR distribution shown in Figures 7–10 and 15–18 for vertical and horizontal polarisations, respectively, the highest field intensity is located near the sharp edges of water container models. The excitation of hydrogen bonding with a total amount of energy higher than 0.238 *eV* by more than one photon is very high in the areas of water located at the sharp edges of water containers.

The concentration of EM energy at the sharp edges is due to the concentration of the amplitudes of electric field at the edges of the containers, and it is a well-known phenomenon in the interaction between the EM waves at microwave frequency and the sharp edges of food products. The energy concentration at the edges is well known in the microwave heating process by 'edge overheating' [6]. In the solid state of materials, such as solid food products, the edge overheating phenomenon is limited to specific areas at the edges of materials, but for materials in the liquid state, such as water, the heating effect will extend to other areas because of the Brownian motion of liquid molecules and particles. For example, the exposure of water containers in real environment to intermittent EM fields will expose a high amount of water to the EM fields that are concentrated at the sharp edges of the containers because of the Brownian motion or the random motion of water molecules and particles. This is why the concentration of EM energy at the sharp edges of water containers plays a very significant role in extending the nonthermal effects of low-intensity EM radiation in water. In addition, the variation in the energy distribution induced in water in the four container models can cause variations in the nonthermal effects of EM radiation in water stored inside the pyramidal, rectangular, square and cylindrical containers. The effect of the sharp edges of water containers as well as the geometrical shapes of water containers on the distribution and the concentration of EM energy could be considered as a possible mechanism of nonthermal effects of low-intensity EM radiation.

## 4. Conclusion

This research was carried out to evaluate the effect of packaging shape, frequency, irradiation geometry, and polarization on the computation of electric and magnetic fields and SAR induced in rectangular, pyramidal, cylindrical, and square container models at 300, 900, and 2,400 MHz and its relationship the non-thermal effects of the electromagnetic radiation. A detailed investigation for exposure to plane wave has been conducted. The evaluation utilized various configurations for the incident field. Results of analyses presented here showed that the maximum electric field and maximum point SAR induced in water in the pyramidal container model was the highest. The order of the effect on total SAR and maximum point SAR is cylindrical < square < rectangular < pyramidal model at 300, 900, and 2,400 MHz for both vertical and horizontal polarizations. It can be concluded that the variation in the packaging shape, frequency, irradiation geometry, and polarization. In addition to the sharp edges of the container models significantly affects the distribution and calculation of electric and magnetic fields and SAR values induced in the stored water, which in turn can cause variations in the non-thermal effects of the electromagnetic radiation in the stored water. Further studies using EM simulation should be conducted to evaluate how specific packaging shapes can affect the behavior of EM fields and the resulting SAR by using different packaging materials.

**Acknowledgements**

The authors would like to thank lectures, colleague and friends for giving valuable advices. We also wish to acknowledge CST Computer Simulation Technology AG for the evaluation license of CST STUDIO SUITE® 2014 and for technical support.